\documentstyle[12pt,axodraw,epsfig]{article}
\renewcommand{\thefootnote}{\fnsymbol{footnote}}

\newcommand{\prepr}[1] {\begin{flushright}  {\bf #1} \end{flushright} \vskip 1.cm}
\newcommand{\titul}[1] {\boldmath \begin{center}{\Large {\bf #1 } } \end{center}
\vskip 0.8cm}

\newcommand{\autor}[1] {\begin{center}  {\bf \lineskip .3cm #1  }
                        \end{center} }

\newcommand{\lugar}[1] {\begin{center}  {\normalsize \bf \it #1   } \end{center}}
\newcounter{muni}

\makeatletter
\def\fmslash{\@ifnextchar[{\fmsl@sh}{\fmsl@sh[0mu]}}
\def\fmsl@sh[#1]#2{%
  \mathchoice
    {\@fmsl@sh\displaystyle{#1}{#2}}%
    {\@fmsl@sh\textstyle{#1}{#2}}%
    {\@fmsl@sh\scriptstyle{#1}{#2}}%
    {\@fmsl@sh\scriptscriptstyle{#1}{#2}}}
\def\@fmsl@sh#1#2#3{\m@th\ooalign{$\hfil#1\mkern#2/\hfil$\crcr$#1#3$}}
\makeatother
\begin{document}
\hbadness=10000
\pagenumbering{arabic}
\begin{titlepage}

\prepr{hep-ph/0306045\\
\hspace{30mm} KIAS--P03035 \\
\hspace{30mm} June 2003}

\begin{center}
\titul{\bf Searching for a very light Higgs boson \\
at the Tevatron}

\autor{A.G. Akeroyd\footnote{akeroyd@kias.re.kr}}
\lugar{Korea Institute for Advanced Study,
207-43 Cheongryangri 2-dong,\\ Dongdaemun-gu,
Seoul 130-722, Republic of Korea}

\end{center}

\vskip2.0cm

\begin{abstract}
\noindent
Light Higgs bosons ($h^0$) with a mass below 60 GeV
may have escaped detection at LEP due to a suppressed
cross--section for $e^+e^-\to Zh^0$. Their discovery is also problematic 
in standard search channels at the Tevatron Run II and LHC.
Such a $h^0$ can arise in the 
two Higgs doublet model (2HDM) and in the Minimal 
Supersymmetric Standard Model (MSSM) with explicit CP violating phases.
We propose the mechanism $p\overline p \to H^\pm h^0$ which offers 
cross--sections of up to 500 fb in the 2HDM, or up to 100 fb in the MSSM. 
The possibility of a large branching ratio for $H^\pm\to h^0W^\pm$
would give rise to the non--standard signature 
$h^0h^0W^\pm$ which might facilitate detection.

\end{abstract}

\vskip1.0cm
{\bf  PACS index :12.60.Fr,14.80.Cp}
\end{titlepage}
\thispagestyle{empty}
\newpage

\pagestyle{plain}
\renewcommand{\thefootnote}{\arabic{footnote} }
\setcounter{footnote}{0}

\section{Introduction}
The quest for Higgs bosons is of utmost importance at high energy
colliders \cite{Carena:2002es}. The Standard Model (SM) 
predicts one neutral Higgs boson ($\phi^0$) for which a
lower limit on its mass $m_{\phi^0}> 114$ GeV has been obtained
by direct searches at LEP2 in the production channel 
$e^+e^-\to \phi^0Z$ \cite{Hagiwara:fs}.
For the lightest CP--even Higgs boson ($h^0$) of the
Minimal Supersymmetric Standard Model (MSSM) the 
analogous mechanism $e^+e^-\to h^0Z$ has
a mixing angle suppression of $\sin^2(\beta-\alpha)$ 
relative to the cross--section for $\phi^0$. However,
this factor is close to 1 in most of the allowed
supersymmetric (SUSY) parameter space, and
even if $\sin^2(\beta-\alpha)$ is suppressed, the complementary channel
$e^+e^-\to h^0A^0 \sim \cos^2(\beta-\alpha)$ can be used.
Combining these two search mechanisms enabled LEP to obtain the 
bound $m_{h^0}> 90$ GeV in the MSSM \cite{Hagiwara:fs}.

Nevertheless, much weaker numerical bounds
apply to the lightest neutral CP--even Higgs boson $h^0$ in
a general (non--supersymmetric) 
two Higgs doublet model (2HDM) \cite{Kalinowski:1988sp}.
Here the factor $\sin^2(\beta-\alpha)$
is a free parameter and merely taking $\sin(\beta-\alpha)<0.1$
reduces the direct search bound to $m_{h^0}> 20$ GeV \cite{Abbiendi:2000ug}. 
Hence a very light $h^0$ ($m_{h^0}<< 100$ GeV) is not excluded by 
direct searches and is also entirely consistent with 
electroweak precision fits \cite{Chankowski:2000an}.
In the MSSM case the above bound $m_{h^0}>90$ GeV
can be weakened in specific regions of parameter space
if explicit CP violating phases are present
in some SUSY parameters. In this scenario the 
CP--even and CP--odd scalar fields mix 
resulting in three mass eigenstates $h_1,h_2,h_3$ which are
now not definite eigenstates of CP \cite{Pilaftsis:1998dd}, 
\cite{Demir:1999hj}. The coupling $h_1ZZ$ can be very suppressed, 
thus debilitating the LEP searches in the channel $e^+e^-\to h_1Z$.
In addition, the complementary search channel $e^+e^-\to h_1h_2$ 
can be rendered ineffective due to the dominance of the
experimentally challenging decay $h_{2}\to h_1h_1$. 
If the CP violating phases are large ($> \pi/3$), and other
SUSY parameters are chosen to enhance the scalar--pseudoscalar
mixing (called the ``CPX scenario'' in \cite{Carena:2002bb}), 
there remains a region $m_{h_1}<60$ GeV, $3 < \tan\beta < 5$ and 
$120< m_{H^\pm} < 130$ GeV which was not covered by LEP and 
will remain elusive in all standard search channels 
at both the Tevatron Run II and LHC \cite{Carena:2002bb}.
A high energy $e^+e^-$ linear collider operating at $\sqrt s=500$ GeV 
would provide much improved coverage of this region via
$e^+e^-\to h_1Z$ and $e^+e^-\to h_1h_2$ \cite{Akeroyd:2001kt}.

In the meantime, it is of interest to seek alternative 
production mechanisms for a very light $h^0$ 
(of the 2HDM) and $h_1$ (of the MSSM) at existing
colliders such as the Tevatron Run II.\footnote{We will not be concerned 
with a very light pseudoscalar $A^0$ which has zero tree--level coupling
$A^0ZZ$ -- see \cite{Larios:2001ma}.}
Recently, diffractive Higgs production, $p\overline p\to p\overline p
h_1$, has been suggested \cite{Cox:2003xp} which can lead to 
sizeable cross--sections
in the MSSM case of order 100 fb for $m_{h_1}< 20$ GeV. 
This mechanism might offer a favourable signal/background ratio, 
but requires suitable proton tagging detectors to be installed. 
In this paper we suggest an additional production process, 
$p\overline p\to H^\pm h_1,H^\pm h^0$, which is unsuppressed in the
above elusive parameter space.
This mechanism was first proposed in \cite{Akeroyd:2003bt} in the context of
light fermiophobic Higgs bosons ($h_f$) 
with enhanced $h_f\to \gamma\gamma$ decays, and in this paper
we consider its application to the above scenarios of a light
$h^0$ and $h_1$.
We will show that $\sigma(p\overline p\to H^\pm h_1,H^\pm h^0$) 
can be comparable in size to the
diffractive production mechanism, and that the branching ratio (BR)
for $H^\pm\to h_1W^\pm,h^0W^\pm$ can be large 
in the parameter space of interest, which would lead to the
non--standard signature of $h^0h^0W^\pm$. Given the importance of finding 
unsuppressed  mechanisms for producing a very light
$h^0$ or $h_1$, the process $p\overline p\to H^\pm h^0 \to h^0h^0W^\pm$
possibly merits a detailed experimental simulation.

In section 2 we introduce the mechanism $p\overline p\to H^\pm h^0(h_1)$, 
section 3 presents our numerical results with conclusions in section 4.

\section{The mechanism $p\overline p\to H^\pm h^0$}
The cross--section for the process $p\overline p\to H^\pm h^0$
\cite{Akeroyd:2003bt} depends on three input parameters: 
$m_{h^0}$, $m_{H^\pm}$ and the coupling $|W^\pm H^\pm h^0|^2$.

\begin{center}
\vspace{-40pt} \hfill \\
\hspace{1cm}
\begin{picture}(200,70)(0,25) % y_2 controls equation position
\Photon(60,25)(118,25){4}{8}
\ArrowLine(10,55)(60,25)
\ArrowLine(60,25)(10,-5)
\DashLine(168,-5)(118,25){3}
\DashLine(118,25)(168,55){3}
\Text(2,55)[]{$u$}
\Text(2,-5)[]{$\overline d$}
\Text(184,-2)[]{$h^0,h_1$}
\Text(179,55)[]{$H^{+}$}
\Text(90,38)[]{$W^{+}$}
\end{picture}
\end{center}
\vspace{1cm}
This mechanism has been disregarded as an effective way of producing 
$h^0$ and/or $H^\pm$ at hadron colliders (e.g. Tevatron).
This view is justified in the MSSM {\it without} explicit 
CP violating phases in the SUSY parameters. A small 
$\sigma(p\overline p\to H^\pm h^0)$ is ensured since
the coupling $|W^\pm H^\pm h^0|^2$ 
$\sim \cos^2(\beta-\alpha)$ is suppressed in the 
MSSM parameter space for $m_{A^0}> m_Z$.
%In addition, with the dominant Higgs boson decays being 
%$h^0\to b\overline b$ and 
%$H^\pm\to \tau^\pm\nu_{\tau}$ (for $m_{H^\pm}< m_t+m_b$), the background is
%expected to be much larger than the signal.
In addition, the phase space suppression is substantial at Tevatron energies
due to the current lower bounds of $m_{h^0}\ge 90$ GeV and 
$m_{H^\pm}\ge 120$ GeV. However, 
$\sigma(p\overline p\to H^\pm h^0)$ is much larger
if the following conditions are fulfilled \cite{Akeroyd:2003bt}:
\begin{itemize}

\item[{(i)}] There is little or no mixing angle suppression
in the coupling $|W^\pm H^\pm h^0|^2$ 

\item[{(ii)}] $h^0$ is very light

\end{itemize}
Both these conditions are satisfied in the MSSM elusive parameter space 
of a light $h_1$ with suppressed coupling $ZZh_1$.
%Condition (i) is a consequence of the very suppressed 
%coupling $ZZh_1$. Condition (ii) ensures that BR$(h_2\to h_1h_1)$ is
%large, which renders ineffective searches in the channel 
%$e^+e^-\to h_1h_2$.
%If the CP violating phase in $\mu A_t$ is large ($> \pi/3$), and other
%SUSY parameters are chosen to enhance the scalar--pseudoscalar
%mixing (called the ``CPX scenario'' in \cite{Carena:2002bb}), 
%there remains a region $m_{h_1}<60$ GeV, $3 < \tan\beta < 5$ and 
%$120< m_{H^\pm} < 130$ GeV which was not covered by LEP and 
%will remain elusive in all standard search channels 
%at both the Tevatron Run II and LHC \cite{Carena:2002bb}.
In a 2HDM the condition for a light, undetected $h^0$ is merely that
the coupling $ZZh^0$ is small. This corresponds to taking small 
values of $\sin^2(\beta-\alpha)$ which maximizes the coupling
$|W^\pm H^\pm h^0|^2$ $\sim  \cos^2(\beta-\alpha)$.
In both these scenarios the processes $p\overline p\to H^\pm h^0$ 
and $p\overline p\to H^\pm h_1$ are not so suppressed, and thus 
this mechanism may offer a chance of probing the problematic
parameter space. 

In \cite{Akeroyd:2003bt}
this mechanism was applied to a special case of the 2HDM
with a light fermiophobic Higgs, $h_f$, which has a large 
BR$(h_f\to \gamma\gamma$) and thus a relatively high
experimental detection efficiency.
It was shown that $p\overline p\to H^\pm h_f$ at Tevatron energies
can offer cross--sections
$> 100$ fb for $m_{H^\pm}< 100$ GeV 
and $m_{h_f}=50$ GeV. In the two scenarios
of interest in this paper the dominant decay of $h^0$ and $h_1$ is 
to $b$ quarks, ($h^0,h_1\to b\overline b$), and assuming
$H^\pm\to \tau^\pm\nu_{\tau}$ the experimental signature of
$b\overline b \tau\nu_{\tau}$ would suffer from
a much larger background than in the fermiophobic case. 
We are not aware of an experimental simulation although this
may become available soon for the case of the related process
$p\overline p\to H^\pm A^0$, where $m_A>90$ GeV was assumed
\cite{Kanemura:2001hz}. 
Hence it is not clear at this stage if $p\overline p\to H^\pm h^0$ 
would be observable above the background even
if $\sigma(p\overline p\to H^\pm h^0$) were sizeable ($>100$ fb). 
Nevertheless, given the need to probe the problematic 
parameter space we believe it is beneficial to give a numerical 
estimation of $\sigma(p\overline p \to H^\pm h^0$) 
in both the above scenarios. Interestingly, we will show that
BR($H^\pm\to h^0W^\pm$) can be large,
which would give rise to the non--standard signature of $h^0h^0W^\pm$
and might ameliorate the signal--background situation.

In order for $p\overline p\to H^\pm h^0$ to be maximized, $H^\pm$ should 
not be too heavy. It is known that the rare decay $b\to s\gamma$
imposes strong  
lower bounds on $m_{H^\pm}$ in the 2HDM (Model II), but 
these are easily avoided in Model I for $\tan\beta>1$ 
\cite{Borzumati:1998tg}. For Model II type fermionic couplings one has
$m_{H^\pm}>200$ GeV \cite{Borzumati:1998tg}, 
which would render $\sigma(p\overline p\to H^\pm h^0$)
small at Tevatron energies. 
A caveat here is that this bound can be weakened to the current
direct search limits ($m_{H^\pm} > 80$ GeV)
if CP violating phases are added in the 2HDM 
\cite{Bowser-Chao:1998yp}, or in a model with more than 2 
Higgs doublets \cite{Borzumati:1998xr},\cite{Grossman:1994jb}.
Given these possibilities, in our analysis for the 2HDM we will 
consider $m_{H^\pm}$ as light as 90 GeV.
In the MSSM it is known that $b\to s\gamma$ constraints depend strongly
on the flavour sector and $120$ GeV $< m_{H^\pm} <130$ GeV (corresponding
to the elusive region) is permissible parameter space.

\section{Numerical Results}
We will show results for $\sigma(p\overline p\to H^\pm h^0$) for a
light $h^0$ in the 2HDM. The angular factor $\cos^2(\beta-\alpha)$ 
arising from the squared coupling $|W^\pm H^\pm h^0|^2$
is close to 1 in the parameter space of interest of 
$\sin^2(\beta-\alpha)<< 1$, and for definiteness we take 
0.97. The cross--section for 
$h_1$ in the problematic parameter space in the MSSM
can be obtained from the 2HDM cross--section by
taking $m_{h_1}$ and $m_{H^\pm}$ to lie in the elusive region
$m_{h_1}\le 60$ GeV, 120 GeV $\le m_{H^\pm}\le 130$ GeV.
We calculate the coupling $|W^\pm H^\pm h_1|^2$ in this region
by using the public code cph.f \cite{pilaftsis:cph}
(which was used in \cite{Carena:2002bb}) 
and find $|W^\pm H^\pm h_1|^2\ge 0.95$, which is consistent
with our choice of 0.97 used above. We sum over the rates for 
$\sigma(p\overline p\to H^+ h^0$) and $\sigma(p\overline p\to H^- h^0$),
and use the Martin-Roberts-Stirling-Thorne parton distribution
functions (MRST2002) from \cite{Martin:2001es}.
In Fig.1 we plot $\sigma(p\overline p\to H^\pm h^0$) as a 
function of $m_{h^0}$, for $10$ GeV$ < m_{h^0} < 90$ GeV.
Note that we are considering smaller $m_{h^0}$ than in
\cite{Akeroyd:2003bt}, which took $m_{h^0}> 50$ GeV.
We plot several curves corresponding to different
values of $m_{H^\pm}$. For the lightest values of the 
Higgs masses ($m_{h^0}=10$ GeV and $m_{H^\pm}=90$ GeV) we obtain 
cross--sections as large as 450 fb. The MSSM result can be read 
off from the curve for $m_{H^\pm}=127$ GeV and 
$10$ GeV $< m_{h_1} < 60$ GeV, which gives cross--sections of 
$100$ fb $\to 60$ fb.
\begin{figure}
\centerline{\protect\hbox{\epsfig{file=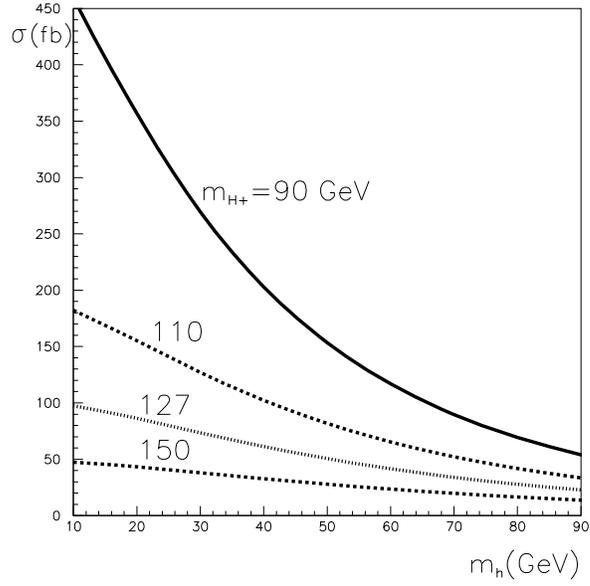,
width=0.50\textwidth,angle=0}}}
\caption{\it Production cross--section $\sigma(p\overline p\to H^\pm h^0$)
at the Tevatron as a function of $m_{h^0}$ for various values of $m_{H^\pm}$,
without including QCD enhancement factor of 1.3.}
\label{cs_all_tb}
\end{figure}
We note that a further enhancement of 
$\sigma(p\overline p\to H^\pm h^0$) comes from the  
the QCD correction factor of 1.3 \cite{Dawson:1998py} which we
have not included in Fig.1.
The cross--sections for the MSSM case are smaller than those for the
diffractive production mechanism \cite{Cox:2003xp} if 
$m_{h_1} < 20$ GeV, but are larger if $m_{h_1}> 20$ GeV.

\begin{figure}
\centerline{\protect\hbox{\epsfig{file=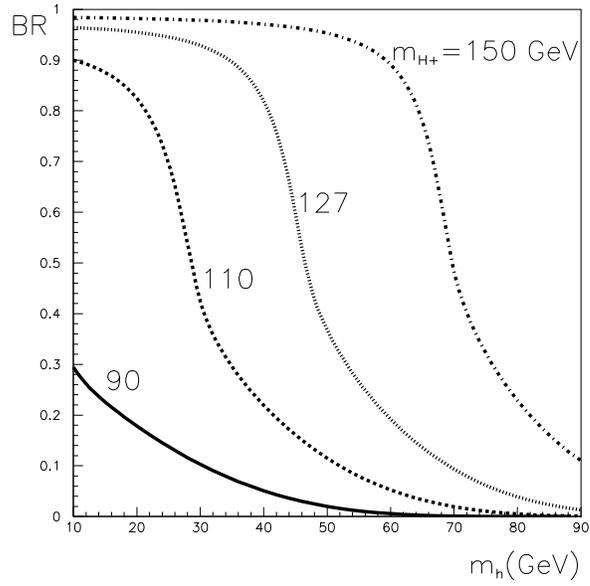,
width=0.50\textwidth,angle=0}}}
\caption{\it BR$(H^\pm\to h^0W^\pm)$ as a function of $m_{h^0}$
for various values of $m_{H^\pm}$.}
\label{cs_all_tb}
\end{figure}

BR$(H^\pm\to h^0W^\pm)$ can be very large when $h^0$ is light. 
Previous analyses of this decay mode have been performed 
in the (CP conserving) MSSM
\cite{Moretti:1994ds}, 2HDM (Model II) \cite{Borzumati:1998xr},
and 2HDM (Model I) \cite{Akeroyd:1998dt}.
For $m_{H^\pm}< m_t+m_b$ of interest to us, the dominant competing 
decay is $H^\pm\to \tau^\pm\nu_{\tau}$
whose rate is proportional to $\tan^2\beta$ for Model II type 
fermionic couplings on which we will focus. In Fig.2 we take 
$\tan\beta=4.2$ for all curves, which is inside the problematic interval
of $3 < \tan\beta < 5$ in the MSSM case. 
%Note that this differs from
%the values $\tan\beta=1,2$ taken in \cite{Borzumati:1998xr}, and
%the decay can proceed on shell which is not permitted in the 
We find BR$(H^\pm\to h^0W^\pm)>80\%$ for the MSSM curve of
$m_{H^\pm}=127$ GeV if $m_{h_1}< 40$ GeV. Thus $H^\pm$ decays
dominantly in this non--standard way in most of the 
elusive parameter space.
In the 2HDM case, the lighter values of $m_{H^\pm}$ (which
have the largest cross--sections)
correspond to smaller, but still sizeable BR$(H^\pm\to h^0W^\pm)$.
This decay would lead to a signature of
$h^0h^0W^\pm$, followed by $h^0\to b\overline b$.
A detailed signal to background simulation would be needed to evaluate 
the detection prospects in this channel.
Given the reasonable cross--sections
we encourage experimental simulations.
We note that some studies of the detection prospects
of $H^\pm\to h^0W^\pm$ decays have been performed 
for LHC energies in the
context of the Next to Minimal Supersymmetric Standard Model
(NMSSM), which may also have a large BR$(H^\pm\to h^0W^\pm)$
\cite{Drees:1999sb}. 
Here the production mechanism $pp\to H^\pm tb$
was used, and promising signal/background ratios were obtained
when BR$(H^\pm\to h^0W^\pm)$ is large.
We are not aware of any such simulations at Tevatron energies.
We note that the analogous mechanism at the Tevatron $p\overline p
\to H^\pm tb$, followed by $H^\pm\to h^0W^\pm$ decay 
could also be used for the scenario of a light $h^0$ and $h_1$.
Here $\sigma(p\overline p
\to H^\pm tb$) is around $50 \to 100$ fb \cite{Carena:2000yx}
for the MSSM elusive region of
120 GeV $< m_{H^\pm} <130$ GeV and $3 < \tan\beta <5$.
These cross--sections are comparable to those of our proposed channel
$p\overline p\to H^\pm h_1$.

\section{Conclusions}
A very light ($< 60$ GeV) Higgs boson $h^0$ would
have escaped detection at LEP if the coupling $h^0ZZ$ were suppressed.
Searching for such a Higgs boson at the Tevatron Run II and the LHC 
in standard channels is also problematic. 
We considered two models which may provide such a $h^0$;
the 2HDM and the MSSM with SUSY sources of CP violation (the 
latter for a very specific parameter choice).
We showed that the mechanism $p\overline p\to H^\pm h^0$ 
at the Tevatron Run II offers 
sizeable cross--sections of up to 500 fb in the 2HDM
and 100 fb in the MSSM. The possibility of a large branching ratio for
$H^\pm\to h^0W^\pm$ would lead to the non--standard 
$h^0h^0W^\pm$ signature. Given the reasonable cross--sections
we encourage experimental simulations of this production 
mechanism for a light $h^0$.

%\section*{Acknowledgements}

\end{document}